\documentstyle[12pt,aasms4,epsfig]{article}
\begin{document}

\title{Minkowski Functionals and Cluster Analysis for CMB Maps. }
\author{D. Novikov$^{1,3}$, Hume A. Feldman$^{1}$ and 
Sergei F. Shandarin$^{1,2}$}
\date{ }
\affil{$^1$Department of Physics \& Astronomy,
University of Kansas, Lawrence, KS 66045\\
$^2$ Theoretical Astrophysics Center, Juliane Maries Vej 30,
DK-2100 Copenhagen, Denmark\\
$^3$ Astro-Space Center of P.N.Lebedev Physical Institute, 
Profsouznaya 84/32, Moscow, Russia.\\
sergei@kusmos.phsx.ukans.edu\\
feldman@ukans.edu}
\begin{abstract}
We suggest novel statistics for the CMB maps that are sensitive to
non-Gaussian features. These statistics are natural generalizations of
the geometrical and topological methods that have been already used in
cosmology such as the cumulative distribution function and genus.  We
compute the distribution functions of the Partial Minkowski Functionals
for the excursion set above or bellow a constant temperature threshold.
Minkowski Functionals are additive and are translationally and
rotationally invariant. Thus, they can be used for patchy and/or
incomplete coverage. The technique is highly efficient computationally
(it requires only $O(N)$ operations, where $N$ is the number of pixels
per one threshold level). Further, the procedure makes it possible to split
large data sets into smaller subsets.  The full advantage of these
statistics can be obtained only on very large data sets. We apply it to
the 4-year DMR COBE data corrected for the Galaxy contamination as an
illustration of the technique.

{\it Subject headings:} cosmic microwave background,
cosmology, theory, observations.

\end{abstract}

\section{Introduction}
Observations of the Cosmic Microwave Background (CMB) provide valuable
information about the early Universe. In addition to the possibility of
measuring the cosmological parameters the CMB data can provide very
important constraints on the type of the seeds that led to the structure
formation (see e.g. \cite{bon-jaf98}).  
Inflationary theories predict Gaussian density perturbations
with nearly scale-invariant spectrum (e.g. \cite{tur97} and references
therein).  The Gaussianity of the density perturbations results in the
Gaussianity of the CMB temperature fluctuations at the surface of last
scattering. Thus, testing the Gaussianity of the CMB fluctuations
becomes a crucial probe of inflation.  On the other hand, if the seeds
for the structure formation are due to topological defects, such as
strings and textures (e.g. \cite{bra98} and references therein) then the
non-Gaussianity of the temperature fluctuations may probe fundamental
physics at high energies.

However, even if the temperature fluctuations were Gaussian at the
surface of the last scattering they may acquire small non-Gaussianity
due to subsequent weak gravitational lensing ( see e.g. \cite{sel96},
\cite{bern97}, \cite{win98}) as well as due to various astrophysical
foregrounds (see e.g. \cite{ban-etal96}).  Higher resolution maps (MAP,
PLANCK) will make it even more problematic.

Establishing the Gaussian nature of the signal is also important for
practical reasons: Some current techniques for estimating the power
spectrum are optimized for the Gaussian fields only (e.g. \cite{fkp},
see also the discussion in \cite{kno-etal98} and \cite{fer-etal98}).

Standard tests for non-Gaussianity are the three-point correlation
function or bispectrum and higher order moments.  However, in practice
negative results of the non-Gaussian tests can hardly be conclusive
since only infinite number of n-point correlation functions can prove
that a field is Gaussian.  A distribution may appear to be Gaussian up
to very high moment and then be non-Gaussian (\cite{ken-stu77}). The
high-order correlation functions are very expensive computationally for
the large data sets ($O(N^m)$, where $m$ is the order of the correlation
function and $N$ is the number of pixels). Thus, any statistic that is
sensitive to non-Gaussianity and computationally efficient is very
useful.

Here are other well known examples of the tests sensitive to
non-Gaussianity: peak statistics (\cite{bon-efs87}, \cite{vit-jus87},
\cite{nov-jor96}),
genus curve (integral geometric characteristics) (\cite{mel89}, 
\cite{col88}, \cite{nas-nov95}); global
Minkowski Functionals (hereafter MFs) (\cite{got-etal90},
\cite{sch-gor98}, \cite{win-kos98}). These functionals also have 
been considered for CMB polarization field by \cite{nas-nov98}. 
Minkowski Functionals
(\cite{min03}) were 'properly' - i.e. in the context of differential and
integral geometry - introduced into cosmology by \cite{mec-etal94} as a
three-dimensional statistics for pointwise distributions in the universe
and then for the isodensity contours of a continuous random field by
\cite{sch-buc97}. We will discuss the MFs in the following section. Here
we just mention that in the two-dimensional case of the temperature maps
the global Minkowski Functionals are the total area of excursion regions
enclosed by the isotemperature contours, total contour length, and the
genus\footnote{In flat space the genus is equal to
the Euler-Poincar\'{e} characteristic.}  or the number of isolated
high-temperature regions minus the number of isolated low-temperature
regions. Partial Minkowski Functionals are the same quantities but used
as characteristics of a single excursion region.

Kogut et al. (1996) measured the 2-point and 3-point correlation
functions and the genus of temperature maxima and minima in the
COBE DMR 4-year sky maps. They concluded that all statistics were in
excellent agreement with the hypothesis of Gaussianity.
\cite{col-etal96} measured the genus of the temperature fluctuations in
the COBE DMR 4-year sky maps and came to a similar conclusion. 
Heavens (1998) computed the bispectrum of the 4-year COBE datasets
and concluded that there was no evidence for non-Gaussian behavior.

However, Ferreira, Magueijo \& G\'{o}rski (1998) studied the
distribution of an estimator for the normalized bispectrum and concluded
that the Gaussianity is ruled out at the confidence level at least of
99\%.
 
The first analysis of two-dimensional theoretical maps of the
temperature fluctuations that used the total area, length of the
boundary and genus for the excursion set was done by \cite{got-etal90}
although without referring to Minkowski Functionals.  Then
\cite{sch-gor98} discussed the application of the MFs to the COBE maps
stressing the importance of taking into account the curvature of the
celestial sphere (the manifold supporting the random field of the
temperature fluctuations).
\cite{sch-gor98} actually applied the statistics to the COBE data  and 
argued for its advantage as a test of non-Gaussian signal. They concluded
that the field is consistent with a Gaussian random field on degree
scale.

\cite{sah-sat-sh98} suggested that the Partial Minkowski Functionals
(PMF) may be used as quantitative descriptors of the geometrical
properties of the elements of the large-scale structure (superclusters
and voids of galaxies).  In particular, they argued that two simple
functions of the PMFs called the shapefinders can distinguish and
reasonably quantify the structures like filaments, ribbons and pancakes.

Here, we describe a new  statistical tool which  is sensitive to non-Gaussian
behavior of random fields. We suggest using the distribution functions of the
Partial Minkowski Functionals and the number of maxima for a given threshold
level. As an illustration we apply it to the
full sky temperature maps obtained by subtracting the galaxy
contributions from 4-year COBE observations (\cite{bennet92}, \cite{bennet94})
The main goal of the paper is the demonstration of the potentiality of this
novel technique.

The outline of the paper is as follows. In Section 2 we briefly review the
Minkowski Functionals. In Sec. 3 we  outline the numerical algorithm 
used for the  evaluation of the Minkowski Functionals. Then, as an example of
application this method  we present the first results of analysis of the
4-year COBE maps. Finally, in Sec. 4 we discuss the results and  
potentiality of the method. 

\section{Minkowski Functionals}

Let us consider one connected region $R_i$ of the excursion set with
$\nu(\theta,\varphi) \equiv (\Delta T(\theta,\varphi)/T)/\sigma_0 >
\nu_t$, where $\sigma_0={\langle (\Delta T/T)^2\rangle}^{1/2}$,
$\nu(\theta,\varphi)$ is the measure and $\nu_t$ is the threshold.  If
the region is complex then it may require very many parameters to fully
characterize it. However, we consider only three parameters: the area of
the region, $a_i$, the length of its contour, $l_i$, and the number of
holes in it $n_{hi}$. These are three partial Minkowski Functionals.  In
the following analysis we also count the number of maxima of
$\nu(\theta,\varphi)$. To obtain the global Minkowski Functionals, we
compute these quantities for all disjoint regions of the excursion set,
i.e. taking the sums $A=\Sigma a_i$, $L=\Sigma l_i$ and $G =\Sigma g_i$: 
``number
of isolated $\nu > \nu_t$ regions'' $-$ ``number of isolated $\nu <
\nu_t$ regions''. The last quantity is the genus introduced into cosmology
long ago by \cite{dor70} and \cite{got-etal86}.  The total area
$A(\nu_t)$ is clearly proportional to the cumulative distribution
function of the random field.  At high (low) threshold levels the excursion
regions appear as isolated hot (cold) spots.

The Minkowski Functionals have several mathematical properties that make
them special among other geometrical quantities.  They are
translationally and rotationally invariant, additive
\footnote{In particular, additivity means that the Minkowski Functionals
of the union of several disjoint regions can be easily obtained if
the Minkowski Functionals of every region is known.}, and have simple
and intuitive geometrical meanings. In addition, it was shown
(\cite{had57}) that all global morphological properties (satisfying
motional invariance and additivity) of any pattern in $D$-dimensional
space can by fully characterized by $D+1$ Minkowski Functionals.

Global MFs of Gaussian fields are known analytically; in the two-dimensional
flat space they are:
\begin{equation}
\begin{array}{l}
A(\nu)=\frac{1}{2}-\frac{1}{2}\Phi(\frac{\nu}{\sqrt{2}}),\\
L(\nu)=\frac{1}{8\theta_c}\nu e^{-\frac{\nu^2}{2}},\\
G(\nu)=\frac{1}{(2\pi)^{3/2}}\frac{1}{2\theta_c^2}\nu 
e^{-\frac{\nu^2}{2}}.
\end{array}
\end{equation}
where $\Phi(x)=\frac{2}{\sqrt{\pi}}\int\limits_{0}^{x}e^{-x'^2}dx'$ is the
error function. Their dependence on the spectrum can be expressed only
in terms of the length scale of the field 
$\theta_c=\frac{\sigma_0}{\sigma_1}$ where $\sigma_0$ and $\sigma_1$
can be calculated from the spectrum $C_l$
\begin{equation}
\begin{array}{l}
\sigma_0=1/4\pi\sum_l (2l+1)C_l,\\
\sigma_1=1/4\pi\sum_l (2l+1)(l+1)lC_l.\\
\end{array}
\end{equation} 

The analytic formulae for the partial Minkowski Functionals are not
known even for Gaussian fields.  However, it does not impose a principal
obstacle in their application since they can be calculated
numerically. It is worth stressing that in practical application, in
addition to the mean value of some quantity one has to know its
variance.  In most cases the variance is not available in an analytic
form even if the mean value is.  For instance, one can calculate
analytically the mean number of hot/cold spots but the variance of this
number can be estimated only numerically.

\section{Application to Two-dimensional Maps}

We identify all disjoint regions above a given threshold $\nu > \nu_t$ for
positive peaks and below the threshold $\nu <- \nu_t$ for negative peaks.
For every region $R_i$ we compute three Minkowski Functionals: 

\noindent The area, $v_{1}^i=a_i$ \\
The perimeter $v_{2}^i=l_i$ i.e. the length of the boundary \\
The number of holes - equivalent to genus - $v_{3}^i=g_{i}$ \\
The number of maxima $v_{4}^i=n_{mi}$ within the region. 

\noindent We then study the cumulative distribution functions
$F(\nu_t,v^k)$ ($ k=1,2,3,4$) of these quantities.

\subsection{Data}

We used the DMR 4--year whole sky maps, where all Galactic emission was
removed. Two independent methods were employed to separate the Galactic
foreground from the cosmic signal. The maps' construction is described
in detail in two papers (\cite{bennet92}, \cite{bennet94}), they were
released in the DMR Analyzed Science Data Sets (ASDS). The two
techniques are: One method is the so--called combination method (map
1). Here they cancel the Galactic emission by making a linear
combination of all DMR maps, then cancel the free--free emission
assuming a free--free spectral index and finally normalize the cosmic
signal in TD temperature. The subtraction method (map 2) constructs
synchrotron and dust emission maps and subtracts them from the DMR data
sets. The Galactic free--free is then removed. The analysis presented in
this paper was done for both maps and the results of the analysis were
similar.

\subsection{Numerical Algorithm}

In this subsection, we describe the numerical algorithm for 
calculation of the distribution of the partial Minkowski 
functionals on the sphere and application of this algorithm to
the COBE data, considered in the previous section.

\subsubsection{Maps simulations.}

In our simulations we use spherical coordinate system to assign pixels
on a sphere. Here we consider the temperature distribution on the
pixelized map as the function of two variables in the coordinate system:
$-\pi/2<\theta <\pi/2$ and $-\pi<\varphi <\pi$. In fact, this function is
defined only in the points $(\theta_{k_1},\varphi_{k_2})$, so that:
\begin{equation}
\nu_{k_1,k_2}=\nu(\theta_{k_1},\varphi_{k_2}), \hspace{0.5cm}\theta_{k_1}=
k_1h_{\theta},\hspace{0.5cm}
\varphi_{k_2}=k_2h_{\varphi}\ .
\end{equation}
We also assume, that $h_{\theta}=h_{\varphi}=h=\frac{2\pi}{M}$ where $M$
is the number of pixels in the $\varphi$ direction. The
total number of pixels is, therefore, $M^2/2$.

The original COBE maps have been
recalculated according this pixelization in the following way:
\begin{equation}
\Delta T_{data}(\theta,\varphi)=B\int\Delta T_{COBE}(\theta',\varphi')
e^{-\frac{\gamma^2}{2\gamma_0^2}}dcos(\theta')d\varphi'
\end{equation}
where $\Delta T_{COBE}$ and $\Delta T_{data}$ are temperatures defined
in the points of COBE cube pixels and in the points defined by the
variables in Eq. (3) above, $\gamma$ is the angle between the pixels,
$\gamma_0=7^0$ is the smoothing angle and B is the normalization.

The temperature fluctuations are completely characterized by the
spectrum coefficients $C_l^m$. Using this description one can write the
following expression for the temperature of the relic radiation:
\begin{equation}
\begin{array}{l}
\Delta T_{data}(\theta,\varphi)=\sum\limits_{l=2}^{\infty}
\sum\limits_{m=-l}^{m=l}C_l^mY_l^m(\theta,\varphi),\\
\nu_{data}(\theta,\varphi)=\frac{\Delta T_{data}(\theta,\varphi)}
{\langle\Delta T_{data}^2\rangle^{1/2}}
\end{array}
\end{equation} 
where $Y_l^m$ are the spherical harmonics.  The summation in Eq.(5) is
from l=2. The term with l=1 is the dipole component. This term has been
removed from the COBE data before analysis because the contribution of
this term in the $\Delta T$ fluctuations can not be separated from the
contribution due to the motion of the observer relative to the
background radiation.

We have simulated 1000 different Gaussian realizations of the
temperature distributions on a sphere so that we can compare the
distribution of partial Minkowski functionals in the observational data
with a random Gaussian field. This was accomplished in the following way:
\begin{equation}
\begin{array}{l}
\Delta T_{g}(\theta,\varphi)=\sum\limits_{l=2}^{\infty}
\sum\limits_{m=-l}^{m=l}a_l^mC_l^{\frac{1}{2}}Y_l^m(\theta,\varphi),\\
\nu_{g}(\theta,\varphi)=\frac{\Delta T_{g}(\theta,\varphi)}
{\langle\Delta T_{g}^2\rangle^{1/2}}
\end{array}
\end{equation}
where $a_l^m$ are independent random Gaussian numbers with zero mean
ie. $\langle a_l^m \rangle =0$ and with unit variances $\langle
(a_l^m) ^2 \rangle =1$.  
The power spectrum $C_l$ in (6) were obtained using $C_l^m$ from (5) as
follows:
\begin{equation}
C_l=\frac{\sum\limits_{m=-l}^{m=l}(C_l^{m})^2}{(2l+1)}
\end{equation}

\subsubsection{Calculation of Partial Minkowski Functionals}

Let us introduce some threshold $\nu_t$ on a pixelized map of the
CMB. Each isolated hot-spot (regions with $\nu >\nu_t$) can be
considered as the cluster characterized by the area, boundary length and 
Euler characteristic or equivalently by genus
(both are directly related to the number of disjoint boundaries). 
For example, the
total area of the map where $\nu>\nu_t$ is the sum of the areas of all
isolated hot-spots, where $\nu>\nu_t$. The global Minkowski Functionals, i.e.
the total area, total boundary length
and total genus can be found by summation of their partial values over
all clusters on the map.  Computing partial Minkowski functionals on
the pixelized map we require that the
algorithm satisfies the following convergence properties:
\begin{equation}
\begin{array}{l}
v_k^i|_p-v_k^i\rightarrow 0,\hspace{0.5cm}as\hspace{0.5cm}
h\rightarrow 0\\
\hspace{2cm}{\rm and}\\
\frac{(v_k^i|_p-v_k^i)}{v_k^i}\sim O(h^z)\hspace{0.5cm}k=1,2
\end{array}
\end{equation}
where $v_k^i|_p$ denotes k-th Minkowski functional of i-th cluster,
calculated on the pixelized map and $v_k^i$ is the exact value of this
functional on the continuous field. In the following analysis we use
linear interpolation, so that $z=1$ for our algorithm.

Pixels $(k_1,k_2)$ inside the regions where $\nu>\nu_t$ satisfy the
condition $\nu_{k_1,k_2}>\nu_t$. We define pixel $(k_1,k_2)$ inside this
region as the inner boundary pixel if the field is below the threshold
level $\nu_t$ at least in one of its four neighbors
$((k_1+1,k_2),(k_1-1,k_2),(k_1,k_2+1), (k_1,k_2-1))$ (eg.
$\nu_{k_1+1,k_2}<\nu_t$) see Fig. 4.  We approximate a smooth boundary
curve by the polygon using linear interpolation of the field
between inner and outer boundary pixels (Fig. 4) and thus finding the
intersection of the boundary curve with the grid lines:
\begin{equation}
\theta_b=k_1h+h\frac{\nu_t-\nu_{k_1,k_2}}
{\nu_{k_1+1,k_2}-\nu_{k_1,k_2}},\hspace{2cm}
\varphi_b=k_2h
\end{equation}
for $\varphi$ grid lines and
\begin{equation}
\theta_b=k_1h,\hspace{2cm}
\varphi_b=k_2h+h\frac{\nu_t-\nu_{k_1,k_2}}
{\nu_{k_1,k_2+1}-\nu_{k_1,k_2}} 
\end{equation}
for $\theta$ grid lines. $\theta_b$ and $\varphi_b$ denote coordinates
of the boundary points $\vec{X}_m=(\theta_b ,\varphi_b$) on the
polygon. This polygon obviously converge to the smooth boundary line as
$h\rightarrow 0$.  

Then, the algorithm of cluster analysis consists of three
steps.
\begin{itemize}
\item{1)} Identifying the boundaries and computing their lengths.\\
First, we search for closed boundary lines of the level $\nu=\nu_t$.
Then, each set of boundary points is ordered by letting $\vec{X}_{m+1}^n$ to
be the nearest boundary point to the point $\vec{X}_m^n$.
The length of a closed boundary line is:
\begin{equation}
l_n=\sum\limits_{m=1}^{m=M_n+1}|\vec{X}_{m+1}^n-\vec{X}_{m}^n| \ ,
\end{equation}
where
$M_n$ is the total number of boundary points in the n-th closed
line of threshold level ($\vec{X}_{M_n+1}^n=\vec{X}_1^n$) and the norm

$|\vec{X}_{m+1}-\vec{X}_{m}|= [(\theta_{m+1}-\theta_{m})^2+ 
\sin^2(\frac{\theta_{m+1}+\theta_{m}}{2})(\phi_{m+1}-\phi_{m})^2]^{1/2}$.

The first point $X_1$ is arbitrary.
Different boundary lines in the map correspond to the
arrays of boundary points ($\vec{X}_m^n$) and inner boundary pixels
($\vec{Y}_m^n$).  The total boundary of an isolated region $\nu >\nu_t$ 
may consist of a number of closed lines (two lines in Fig.4).

\item{2)} Finding all the boundaries of a connected region (cluster),
computing the total boundary length and genus.\\
We combine all closed lines which are the boundaries of the same cluster
by using arrays of inner
boundary pixels ($\vec{Y}_m^n$). Suppose, we wish to check whether
two different lines are the boundaries of the same
cluster or not. These lines correspond to two sets of inner boundary pixels
$\vec{Y}_m^{n_1}$ and $\vec{Y}_m^{n_2}$. If we take two arbitrary inner
pixels one from each set and connect them by 
a path along grid lines (see Fig. 4) then the path can intersect 
the boundaries $N_{int}^i$ times ($i=1,2$), where $N_{int}^i \ge 0$. If the
both numbers $N_{int}^1$ and $N_{int}^2$ are even then both 
inner boundary pixels
belong to the same cluster otherwise they belong to two
different regions (clusters). Therefore all boundary lines which belong to
one cluster form its boundary with the total boundary length equal 
to the sum of their lengths. 
The number of closed lines for each cluster is equivalent to
the genus of this cluster. Thus we find total number of clusters and two
partial Minkowski functionals for each of them -- the length and genus.

\item{3)} Computation of the areas of clusters.\\
All pixels situated between the inner boundary pixels of a cluster 
belong to the
cluster. The area of the cluster can be roughly approximated 
by the total area of all these
pixels (including inner boundary pixels). 
\end{itemize}

\subsection{Results} 

Figures 5 and 6 show the cumulative distribution functions
$F(\nu_t,V_k)$ (where $V_k=\sum_i v_{k}^i$) for the two COBE maps
described above along with the mean Gaussian values and variances.  The
mean and variance were obtained from 1000 random realizations of
Gaussian fields having the same amplitudes as shown but different sets
of random phases.  In both Fig. 5 and 6 we see significant deviations from
Gaussianity. It may be interesting to note that each statistic shows
the greatest difference with the Gaussian value at different thresholds:
$F(A)$ at $\nu_t= -0.5$, $F(L)$ at $\nu_t= -1$, $F(G)$ at $\nu_t= \pm
1$, and $F(N_m)$ at $\nu_t = 0, 0.5, -1$.  These differences are roughly
the same for both maps and suggest that each of the four statistics carry
different statistical information.

As one might expect, more detailed information can be obtained from the
partial Minkowski Functionals. Figures 7 through 11  show the partial
Minkowski Functionals at ten thresholds $\nu_t= \pm 2,\pm 1.5, \pm 1,
\pm 0.5$ and $\pm 0$\footnote
{The thresholds $\nu_t=+0$ and $-0$ correspond to the
excursion sets $\nu>0$ and $\nu<0$ respectively.}.  Every figure shows
two curves, one for the positive ($\nu>\nu_t $, solid lines) and one for the
negative ($\nu<\nu_t $, dashed lines) thresholds, having same absolute
magnitude $\nu_t $ for each map. Thick and thin
lines correspond to COBE maps 1 and 2 respectively.
The mean Gaussian curve (that obviously
does not depend on the sign of the threshold) is the dotted line and
$1\sigma$ Gaussian variance is shown as a shaded area.

The main features of Fig. 7-11 are the following:
\begin{itemize}
\item Figure 7, $\nu_t=2$: The functions $F(a)$ and $F(l)$ show strong
non-Gaussian signal; $F(g)$ and $F(n_m)$ roughly consistent with the
Gaussianity.
\item Figure 8, $\nu_t=1.5$: All  statistics shows strong non-Gaussianity.
\item Figure 9, $\nu_t=1.$: The strongest non-Gaussian signal comes from the 
distribution of maxima, $F(n_m)$; other statistics are roughly compatible 
with Gaussianity.
\item Figure 10, $\nu_t=0.5$: All statistics show marginal disagreement 
with Gaussianity.
\item Figure 11, $\nu_t=0$: All statistics are in rough agreement with  
Gaussianity.
\end{itemize}

\section{Discussion}

We suggest new statistics to test the Gaussianity of CMB maps.
These statistics are the distribution functions of the Partial Minkowski
Functionals of the excursion set for a given threshold level $\nu_t$. We
also compute the distribution function of the number of maxima in the
isolated region.  The Partial Minkowski Functionals have transparent
geometrical and topological meanings of : 1) the area, 2) the perimeter
and 3)the genus of each disjoint region. Introducing these statistics is
a natural step further. It generalizes the statistical techniques used
before: the cumulative distribution function that correspond to the
total area of the excursion sets, global genus (\cite{dor70},
\cite{got-etal86}), the total length of the boundary
(\cite{got-etal90}). The set of three characteristics mentioned above is
known as global Minkowski Functional and has already been used in
cosmology (\cite{mec-etal94}, \cite{sch-buc97}, \cite{win-kos98},
\cite{sch-gor98}).  \cite{sah-sat-sh98} argued that the Minkowski
Functionals can describe the morphological characteristics of isolated
regions or objects.

It is well known that Minkowski Functionals are invariant under
translations and rotations and are also additive.  Due to these
properties the MFs can be used in cases of incomplete or patchy
coverage. The additivity allows to analyze large data set by splitting
them into a set of smaller subsets.  Measuring the partial MF is very
efficient computationally: for a fixed threshold it requires only $O(N)$
operations, where $N$ is the number of pixels.

Clearly, computing the distribution functions of the partial MF require
much larger data sets than we used here for an illustrative purpose.
However, the upcoming MAP and Plank missions will provide the necessary
resolution. Needless to say that by knowing the partial MFs one can
easily obtain global MFs. Thus, these statistics completely incorporate
the global MFs.

We show that the reconstructions of the whole sky maps obtained by
subtraction the Galaxy contributions from the COBE maps are strongly
non-Gaussian.  This was suspected by many authors who used only the
polar caps for the analysis of the cosmological signal
(e.g. \cite{col-etal96}, \cite{sch-gor98}, \cite{fer-etal98}).  It is
possible that some of the non-Gaussian signal in our analysis is due
to errors in the Galaxy removal, since the Galactic signal is obviously
non-Gaussian. For instance, \cite{col-etal96} found no
deviations from Gaussianity in the genus curve while we see significant
non-Gaussianity in our genus measurements. Unfortunately, by analyzing the
Galaxy caps only we use only about half the data and in addition
split the remaining half into two disjoint regions affected by the
boundaries. Since the largest number of disjoint regions is
about ten at $\nu_t=1.5, 1$ (Fig. 9,10) in the whole sky, the analysis of
the caps only does not seem to be reasonable. This paper provides a
feasibility study in the strengths and usefulness of the Partial Minkowski
Functionals, not a definitive results regarding the non-Gaussian nature
of the COBE--DMR maps. 

Acknowledgments. We are grateful to Ned Wright for criticism and useful
comments.
S. Shandarin thanks K. G\'{o}rski, V. Sahni, B. Sathyaprakash and
S. Winitzki for useful discussions of the related topics and comments.
We acknowledge the support of EPSCoR 1998 grant. DN was
supported by NSF-NATO grant 6016-0703, HF was supported in part by the
NSF-EPSCoR program and the GRF at the University of Kansas. S. Shandarin also
acknowledges the support from GRF grant at the University of Kansas and
from TAC Copenhagen.

\newpage
\begin{figure}
\centerline{\psfig{file=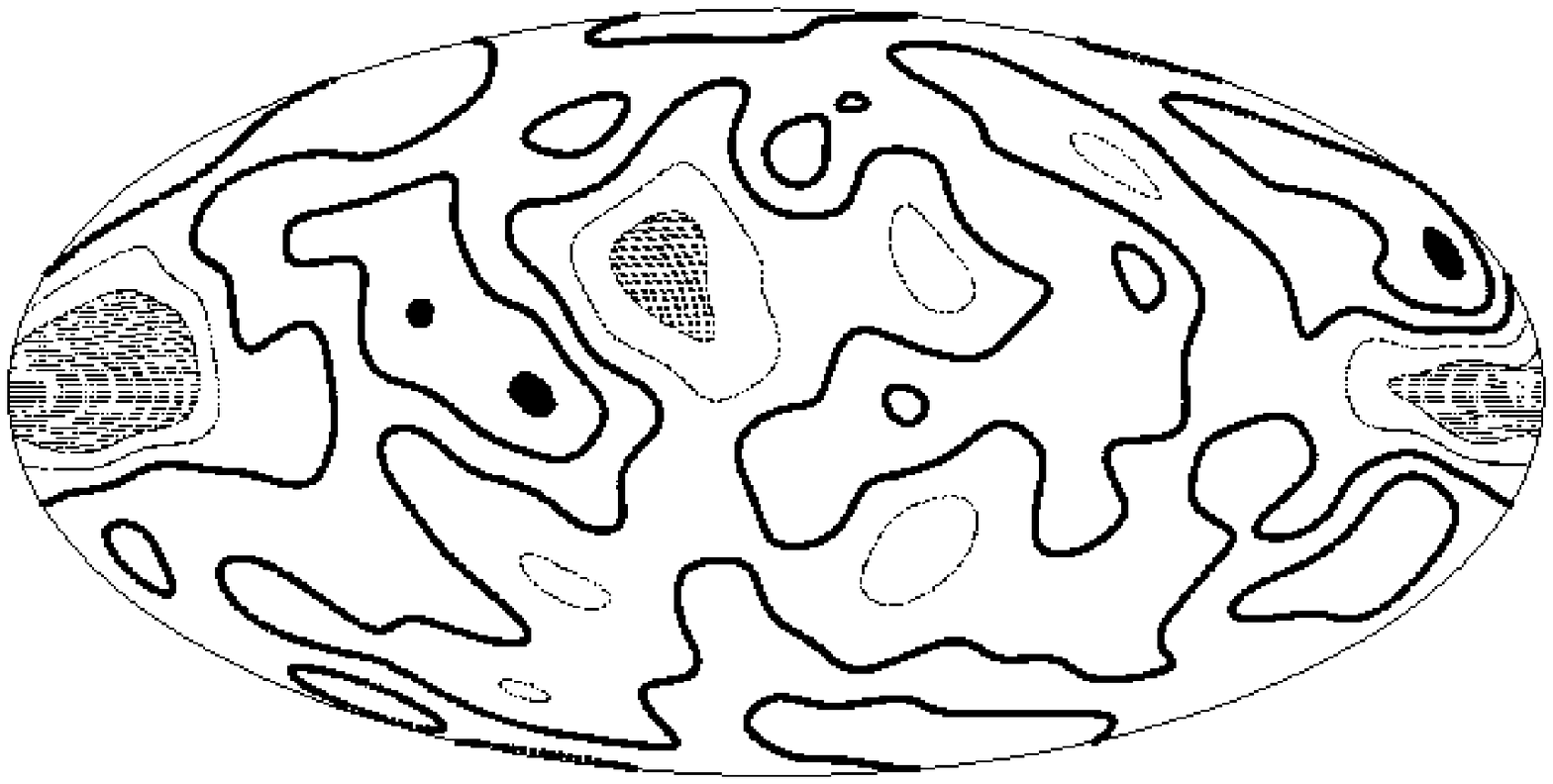,width=12cm}}
\figcaption{The COBE map 1 constructed using the combination
method (see \S3.1). The heavy lines show $0$, $1 \sigma$ and $2 \sigma$ (the
interior of $2 \sigma$ contours is shown in black); the light lines show
$-1 \sigma$ and $-2 \sigma$ (the interior of $-2
\sigma$ contours is shaded).  }
\label{fig1}
\end{figure}
\begin{figure}
\centerline{\psfig{file=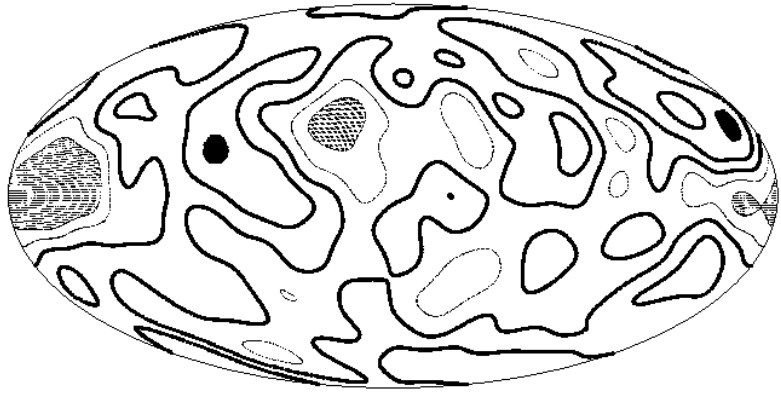,width=12cm}}
\figcaption{The COBE map 2 constructed using the subtraction
method (see \S3.1). The notations are same as in Fig. 1
}
\label{fig2}
\end{figure}
\begin{figure}
\centerline{\psfig{file=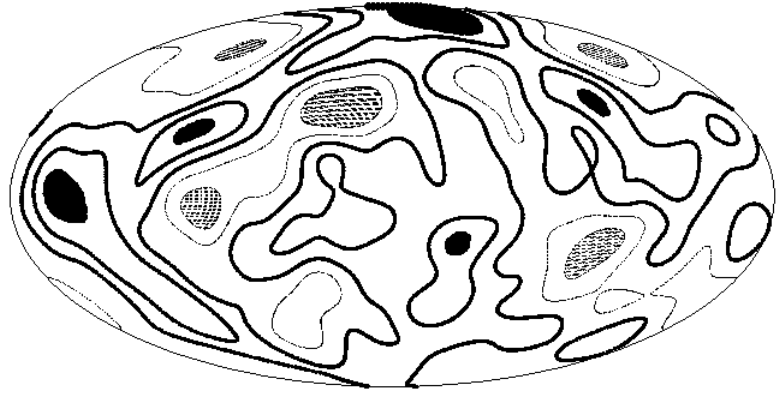,width=12cm}}
\figcaption{An example of a Gaussian map with the same as 
in COBE map 1 amplitudes $l(l+1)C_l/\sigma_0$
}
\label{fig3}
\end{figure}
\begin{figure}
\centerline{\psfig{file=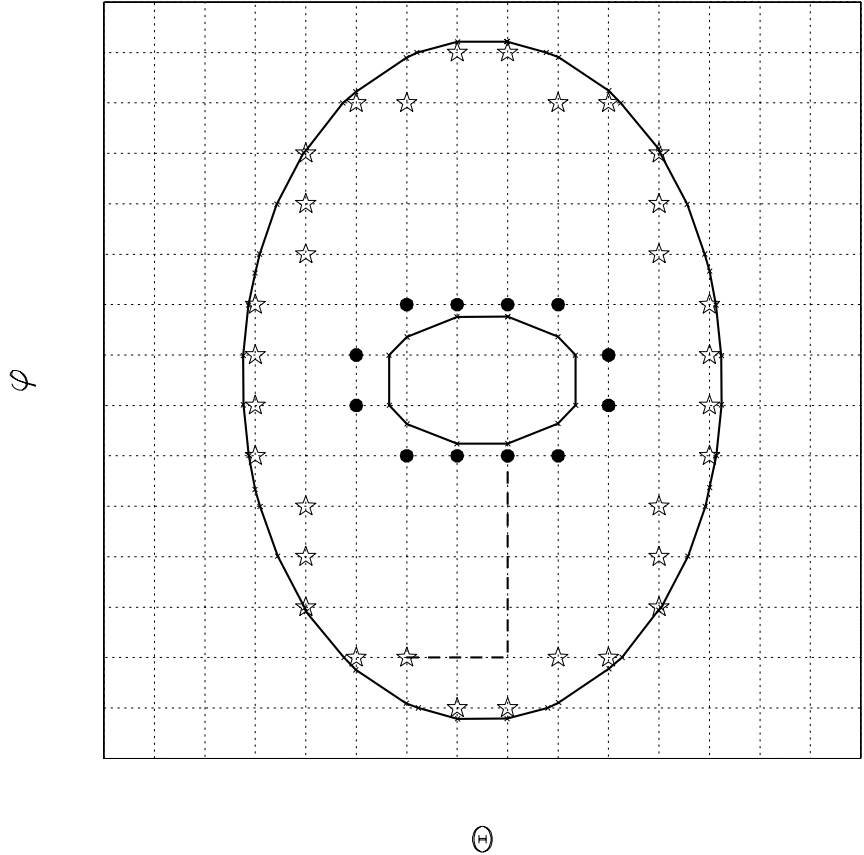,width=12cm}}
\figcaption{A region bounded by the level contours $\nu>\nu_t$ is shown.
The closed polygons are the approximations to the boundaries based on the
linear interpolation.
Stars and circles are two sets of the inner boundary pixels corresponding
to the boundaries. The dashed line shows a possible path on the grid
connecting a pair of the inner boundary pixels that may belong to the
same region or two different regions. If such a path intersects
the both boundaries (i.e. the boundaries corresponding to the pixels
in question)
an even number of times then both pixels belong to
the same (connected) region, otherwise they belong to  different regions.
In this case both numbers are zeros (i.e. even) and thus they belong to the
same region.}
\label{fig4}
\end{figure}
\begin{figure}
\centerline{\psfig{file=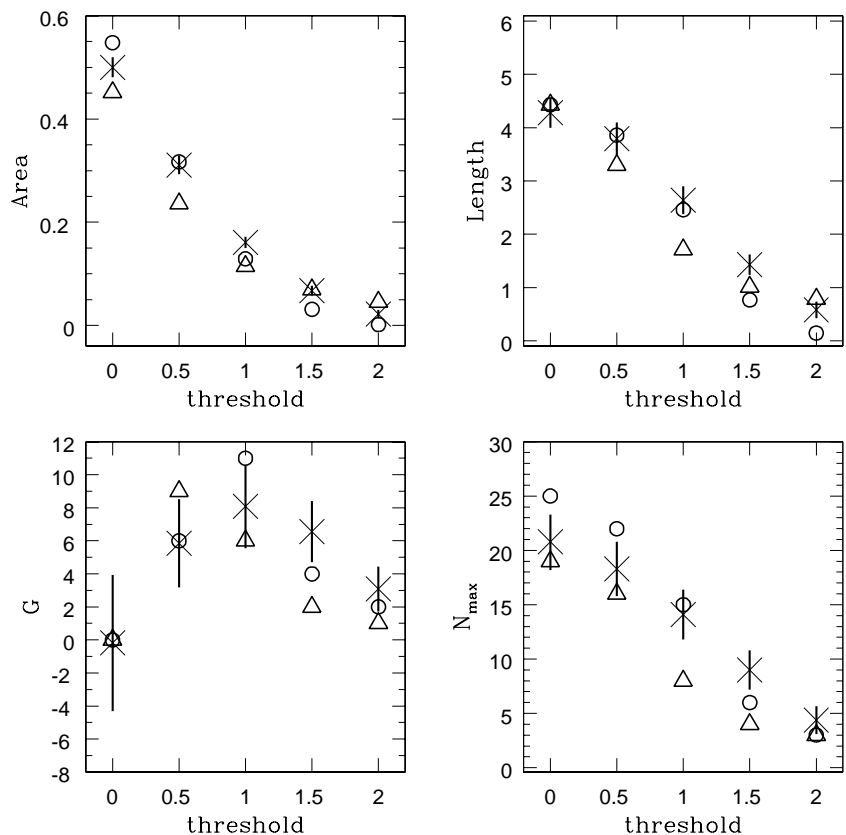,width=12cm}}
\figcaption{The cumulative distribution functions of global 
Minkowski Functionals and the numbers of maxima/minima as a function of
the temperature threshold (given in units of $\sigma$ for COBE map 1.
Circles and triangles show the values for positive and negative
thresholds respectively. The error bars corresponds one $\sigma$
dispersions calculated in 1000 Gaussian realizations.}
\label{fig5}
\end{figure}
\begin{figure}
\centerline{\psfig{file=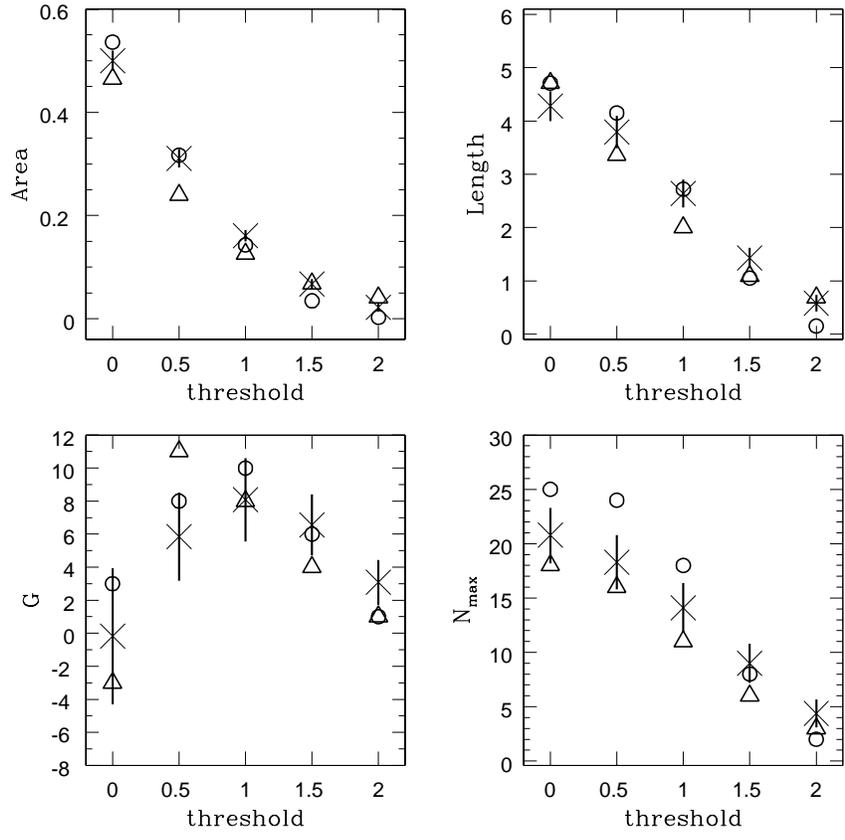,width=12cm}}
\figcaption{ Same as Fig. 5 but for the COBE map 2.}
\label{fig6}
\end{figure}
\begin{figure}
\centerline{\psfig{file=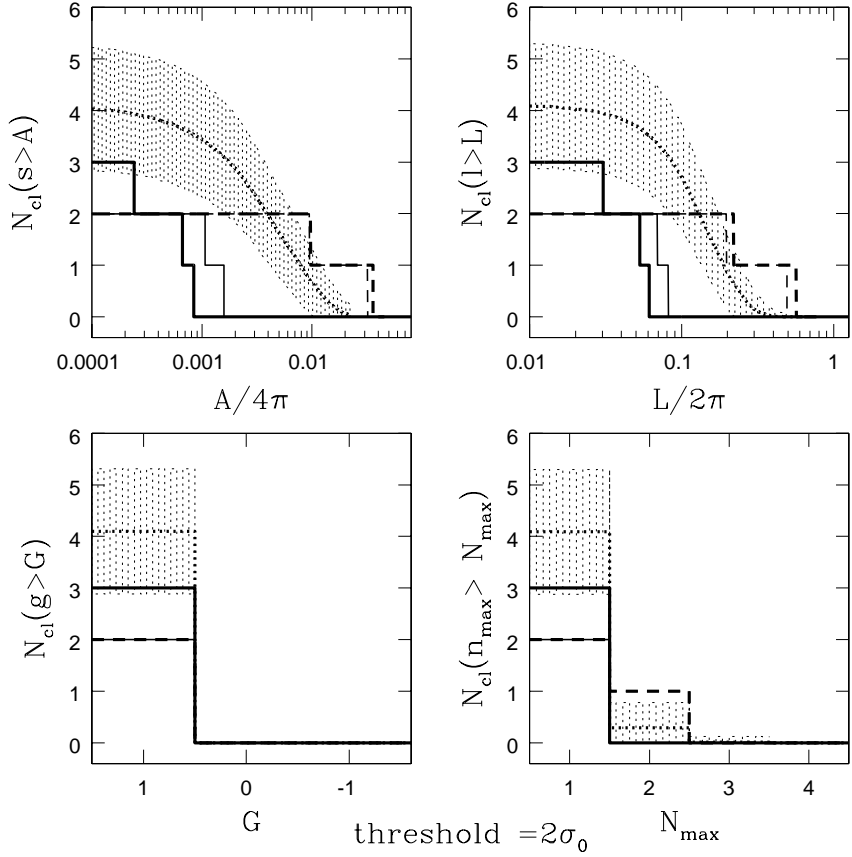,width=12cm}}
\figcaption{The cumulative distribution functions of partial 
Minkowski Functionals for both COBE maps. The shaded strips show
$\pm \sigma$ regions for Gaussian realizations.  Solid line shows the
cumulative distribution functions for positive ($\nu>\nu_t$) and dashed
line for negative ($\nu<\nu_t$) thresholds.  Thick lines correspond to
COBE map 1 and thin lines to COBE map 2.  The threshold
$\nu_t=2\sigma$. }
\label{fig7}
\end{figure}
\begin{figure}
\centerline{\psfig{file=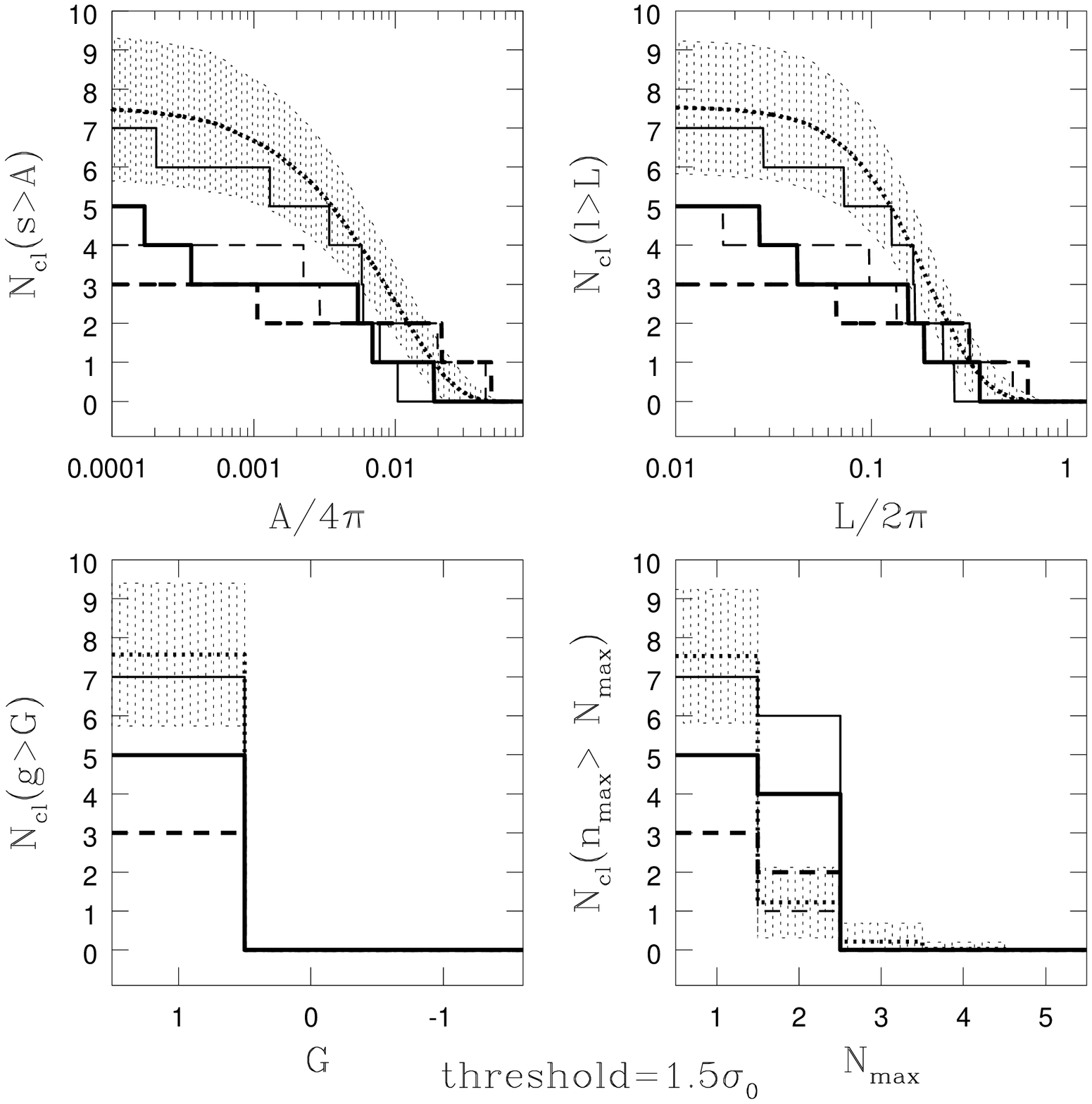,width=12cm}}
\figcaption{Same as Fig. 7 with $\nu_t=1.5\sigma$.}
\label{fig8}
\end{figure}
\begin{figure}
\centerline{\psfig{file=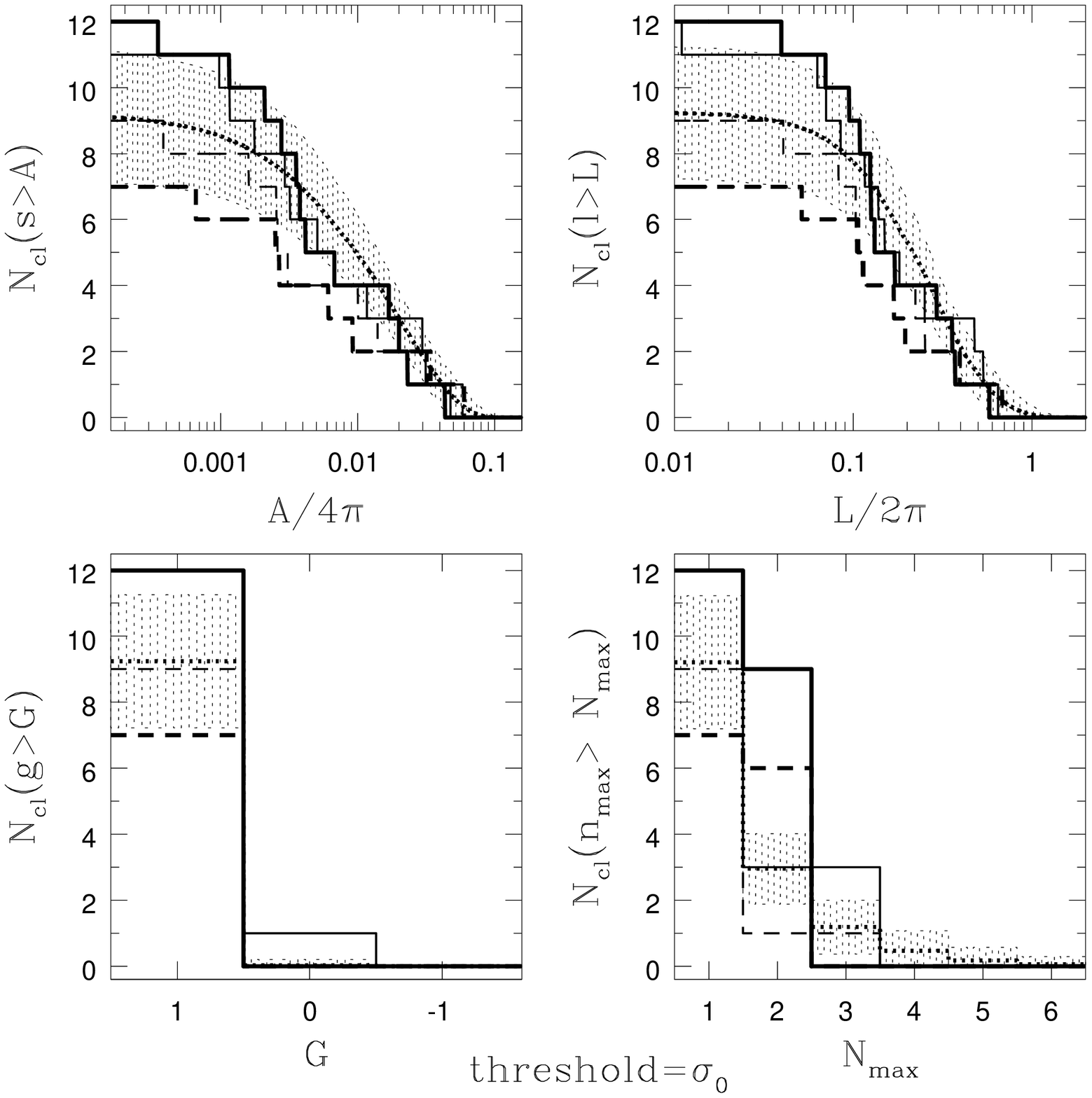,width=12cm}}
\figcaption{Same as Fig .7 with $\nu_t=1\sigma$.}
\label{fig9}
\end{figure}
\begin{figure}
\centerline{\psfig{file=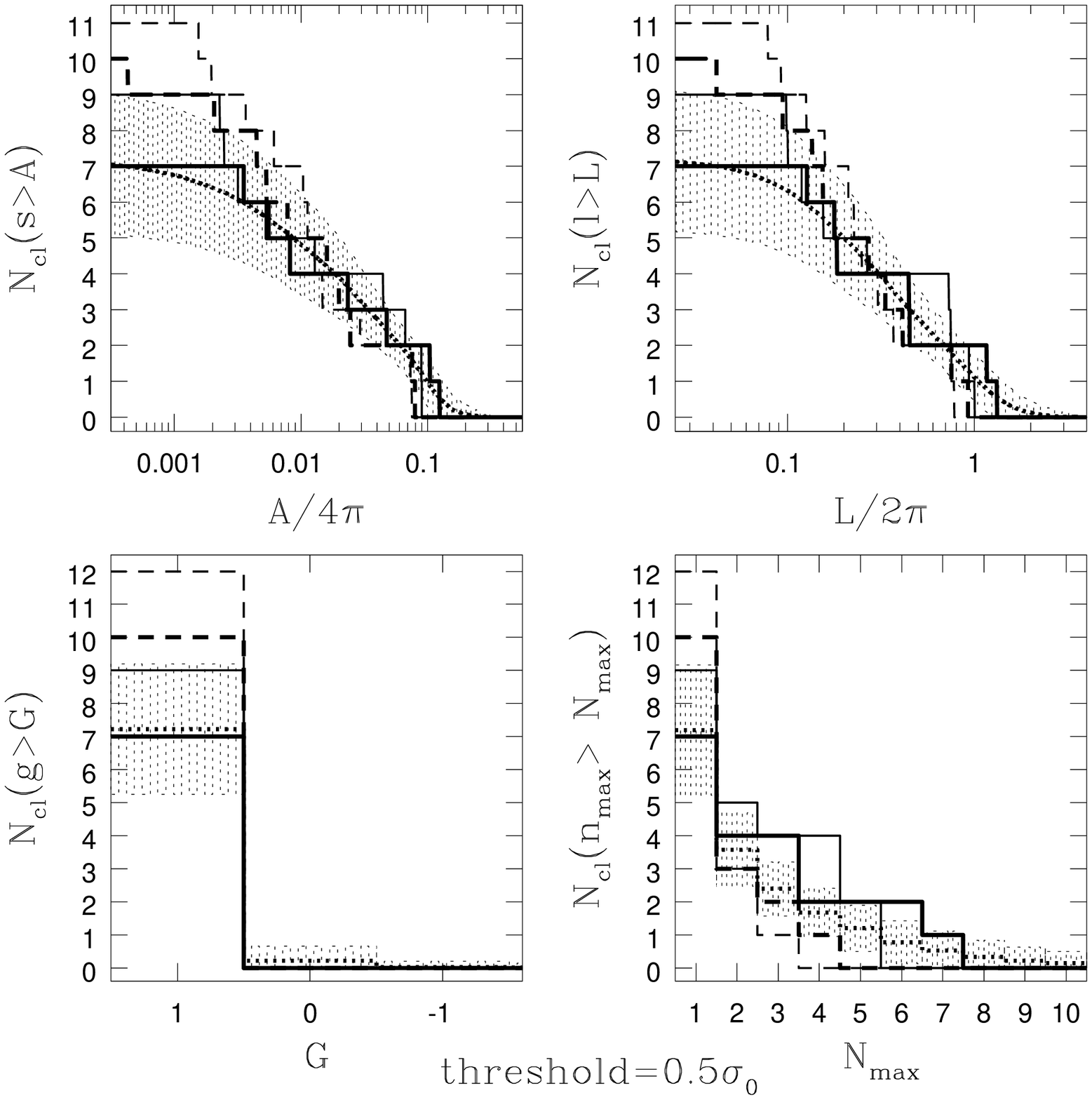,width=12cm}}
\figcaption{Same as Fig. 7 with $\nu_t=0.5\sigma$.}
\label{fig10}
\end{figure}
\begin{figure}
\centerline{\psfig{file=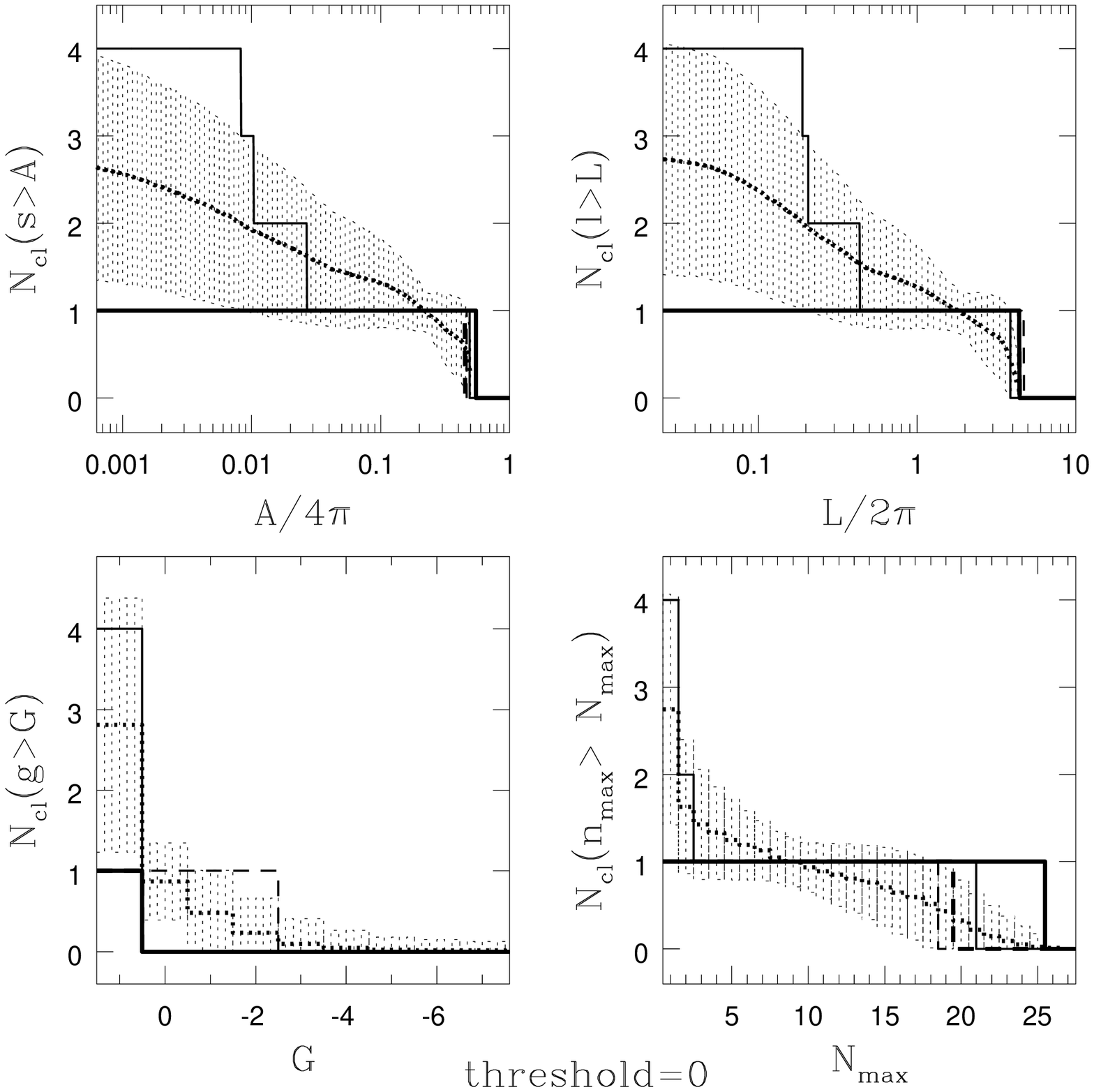,width=12cm}}
\label{fig11}
\figcaption{Same as Fig. 7 with $\nu_t=0$.}
\end{figure}

\newpage
\begin{thebibliography}{}
\bibitem[Bandy et al 1996]{ban-etal96}
Bandy, A.J., G\'{o}rski, K.M., Bennett, C.L., Hinshaw, G., Kogut, A., \&
Smoot, G.F. 1996, \apj, 468, L85
\bibitem[Bennet et al. 1994]{bennet94}
Bennett, C.L. et al., 1994, \apj, 436, 423
\bibitem[Bennet et al 1992]{bennet92}
Bennett, C. L., et al. 1992, \apj, 396, L7-L12
\bibitem[Bernardeau 1997]{bern97}
Bernardeau, F. 1997, \aap, 324, 1
\bibitem[Bond \& Efstathiou 1987]{bon-efs87}
Bond J.R., Efstathiou,  G. 1987, \mnras, 226, 655
\bibitem[Bond \& Jaffe 1998]{bon-jaf98}
Bond, J.R., Jaffe, A.H. 1998, astro-ph/9809043
\bibitem[Brandenberger 1998]{bra98}
Brandenberger, R.H., 1998, astro-ph/9806473
\bibitem[Coles 1988]{col88}
Coles, P. 1988, \mnras, 234, 509
\bibitem[Colley et al 1996]{col-etal96}
Colley, W.N., Gott, III, J.R., \& Park, C. 1996, \mnras, 281, L82 
\bibitem[Doroshkevich 1970]{dor70}
Doroshkevich, A.G. 1970, Astrophysics, 6, 320 
\bibitem[Feldman et al 1994]{fkp}
Feldman, H.A., Kaiser, N. \& Peacock, J. 1994,
\apj, 426 23--37
\bibitem[Ferreira et al 1997]{fer-etal97}
Ferreira, P.G., Magueijo, J., \& Silk, J. 1997 Phys.Rev., D56, 7493
\bibitem[Ferreira et al 1998]{fer-etal98}
Ferreira, P.G., Magueijo, J., \& G\'{o}rski, K.M. 1998, ApJ, 503, L1
\bibitem[Gott et al 1990]{got-etal86}
Gott, III, J.R., Melott, A.L., \& Dickinson, M. 1986, \apj, 306, 341
\bibitem[Gott et al 1990]{got-etal90}
Gott, III, J.R., Park, C., Juskiewicz, R., Bies, W.E., Bennett, D.P., 
Bouchet, F.R., Stebins, A. 1990, \apj, 352, 1
\bibitem[Heavens 1998]{hea-98}
Heavens, A.F. 1998, MNRAS, 299, 805
\bibitem[Hadwiger 1957]{had57}
Hadwiger, H. 1957, Vorlesungen \"{u}ber Inhalt, Oberfl\"{a}che und
Isoperimetre, Springer-Verlag, Berlin)
\bibitem[Kendall \& Stuart 1977]{ken-stu77}
Kendall, M.G. \& Stuart, A. 1977, , The Advanced Theory of Statistics,
4th edition, Charles Griffin.
\bibitem[Knox et al 1998]{kno-etal98}
Knox, L., Bond, J.R., Haffe, A.H., Segal, M., Charbonneau, D. 1998,
Phys.Rev. D58, 083004
\bibitem[Kogut et al 1996]{kog-etal96}
Kogut, A., Bandy, A.J., Bennett, C.L., G\'{o}rski, K.M., Hinshaw, G., 
Smoot, G.F., \& Wright, E.L. 1996, \apj, 464, L29
\bibitem[Mecke, Buchert, \& Wagner 1994]{mec-etal94}
Mecke, K.R., Buchert, T, Wagner, H. 1994, \aap, 288, 697
\bibitem[Melott et al 1989]{mel89}
Melott, A., Cohen, A., Hamilton, A., Gott, J., \&
Weinberg, D. 1989, \apj 345 618
\bibitem[Minkowski 1903]{min03}
Minkowski, H. 1903, Math. Ann., 57, 447
\bibitem[Naselsky \& Novikov 1995]{nas-nov95}
Naselsky \& Novikov 1995, \apj, 444, L1 
\bibitem[Naselsky \& Novikov 1998]{nas-nov98}
Naselsky \& Novikov 1998, \apj, 507, 31
\bibitem[Novikov \& Jorgensen 1996]{nov-jor96}
Novikov \& Jorgensen 1996, \apj 471 521
\bibitem[Sahni, Sathyaprakash \& Shandarin 1998]{sah-sat-sh98}
Sahni, V., Sathyaprakash, B.S., Shandarin, S.F. 1998, \apj, 495, L5
\bibitem[Schmalzing \& Buchert 1997]{sch-buc97}
Schmalzing, J. \& Buchert, T. 1997, \apj, 482, L1
\bibitem[Schmalzing \& G\'{o}rski 1998]{sch-gor98}
Schmalzing, J., G\'{o}rski, K.M. 1998, \mnras, 297, 355
\bibitem[Seljak 1996]{sel96}
Seljak, U. 1996, \apj, 463, 1
\bibitem[Turner 1997]{tur97}
Turner, M. 1997, in ``Generation of Cosmological Large-Scale Structures.''
eds. D.N. Schramm and P. Galeotti, Kluwer Academic Publishers, p. 153
\bibitem[Vittorio \& Juskiewicz 1987]{vit-jus87}
Vittorio, N. \& Juskiewicz, R. 1987, \apj, 314, L29
\bibitem[Winitzki 1998]{win98}
Winitzki, S. 1998, astro-ph/9806105
\bibitem[Winitzki \& Kosowsky 1998]{win-kos98}
Winitzki, S. \& Kosowsky, A. 1998, New Astronomy, 3, 75
\end{thebibliography}
\end{document}